\documentstyle[a4,12pt,epsfig]{article}
\newcommand{\beq}{\begin{equation}}
\newcommand{\eeq}{\end{equation}}
\newcommand{\bea}{\begin{eqnarray}}
\newcommand{\eea}{\end{eqnarray}}

\newcommand{\aas}{/\kern-.50em A}

\begin{document}

\title{{\bf Bogomol'nyi equations in gauge theories}}
\author{M.S. Cunha,\ \ H.R. Christiansen \\
{\normalsize {\it Centro Brasileiro de Pesquisas
F\'{\i}sicas}}\thanks{ Electronic address: marcony@cat.cbpf.br,
hugo@cat.cbpf.br}\\ {\normalsize {\it Fields and Particles
Department, CBPF - DCP}}\\ {\normalsize {\it Rua Dr. Xavier Sigaud
150, 22290-180}}\\ {\normalsize {\it Rio de Janeiro, Brazil.}}\\
{\normalsize and} \\ C.A.S. Almeida \\ {\normalsize {\it
Universidade Federal do Cear\'{a}}}\, \thanks{
carlos@fisica.ufc.br}\\ {\normalsize {\it Physics Department}}\\
{\normalsize {\it C.P. 6030, 60455-760 Fortaleza-Ce, Brazil}}}
\maketitle

\begin{abstract}
By imposing self-duality conditions, we obtain the explicit form in which
gauge theories spontaneously breakdown in the Bogomol'nyi limit. In this
context, we reconsider the Abelian Higgs, Chern-Simons Higgs and
Maxwell-Chern-Simons Higgs models. On the same footing, we find a
topological Higgs potential for a Maxwell-Chern-Simons extended theory
presenting both minimal and nonminimal coupling. Finally, we perform a
numerical calculation in the asymmetric phase and show the solutions to the
self-dual equations of motion of the topological theory. We argue about
certain relations among the parameters of the model in order to obtain such
vortex configurations.
\end{abstract}

\section{Introduction}

The study of classical vortex solutions in gauge theories, extensively
developed in the seventies, has motivated many works in the last few years.
Actually, their existence has been claimed much earlier in connection to the
Ginzburg-Landau model of superconductivity, in the pioneering work of
Abrikosov \cite{Abrikosov}; later on, such topological configurations were
experimentally observed in type-II superconductors \cite{type-II}.

In a relativistic framework, the issue of soliton solutions has been
generalized to field theories such as the Abelian Higgs model in which
vortices exhibit nonzero magnetic fluxes, but are electrically neutral \cite
{Nielsen-Olesen}. By adding a Chern-Simons term, interesting features arise.
In particular, vortex solutions gain an electric charge in the so-called
topologically massive electrodynamics \cite{Paul-Khare,Schapo}.

Self-duality, which allows the reduction of the second order equations of
motion to a set of first order ones, is an interesting tool from the
physical point of view since it is related to the minimization of the energy
together with the action of the system \cite{Dunne}.

Gauge fields whose dynamics exclusively depend on the Chern-Simons term and
are minimally coupled to the scalar fields, have been thoroughly studied by
Jackiw {\it et al.} in recent years \cite{JW,JLW}. Such models present
topological and nontopological solitons and the Bogomol'nyi equations are
obtained for specific sixth-order Higgs potentials.

On the other hand, when gauge field dynamics are also controlled by
a Maxwell term, it seems necessary to introduce a neutral scalar
field in order to ensure self-dual solutions to the model
\cite{LLM}. In fact, this can be avoided by appealing to an extra
{\it nonminimal} contribution to the covariant derivative, which
can be interpreted as an anomalous magnetic moment \cite{Stern}. It
is worth noting however that this analysis has been generally
limited to nontopological potentials presenting just a symmetric
phase \cite{Torres,LeeMin}.

The purpose of this paper is twofold. In a first part we review some
well-known gauge theories presenting vortices and self-dual solutions.
Though, in contrast to preceding authors, we work along the lines given in
ref.\cite{Prem} where self-duality conditions supply the well-known Higgs
potentials for the Abelian Higgs and Chern-Simons Higgs models. Our aim at
this point is to compare this procedure to the standard one and extract the
Bogomol'nyi equations from the energy functional rather than imposing them
together with the equations of motion as in \cite{Prem}.

In a second part, we go a step further and apply this idea to set
the proper Bogomol'nyi conditions and obtain the resulting Higgs
potential in a Maxwell-Chern-Simons Higgs model coupled
nonminimally as above mentioned. This kind of coupling has been
typically confined to a critical value that provides the model
fractionary statistic solutions proper of pure CS theory. In this
respect, let us stress that we work without any specific choice of
the anomalous magnetic moment coupling, its value being only
conditioned by the basic assumption of positive energy solutions.
Neither we impose rotational symmetry before taking the Bogomol'nyi
limit, which could hide true minimal energy solutions (see e.g.
\cite{LeeMin}). This freedom allows a tuning between Maxwell and
Chern-Simons contributions, as will be shown later in the series
expansion analysis of the potential. In this way we are able to
find a {\it topological} potential for this model, without
restricting us to a critical nonminimal coupling ({\it c.f.}
\cite{Torres} where only a nontopological sector is found in such a
critical regime)). Finally, we perform a numerical calculation of
the solutions to this theory for a convenient ansatz.

Our results put forward certain relations among the parameters of the model
which, in particular, exclude the usual choice for the topological mass
constants, see refs.\cite{LeeMin,Torres}.


\section{Minimal Models}

\subsection*{Abelian Higgs Model}

Let us start with a Higgs model Lagrangian
\begin{equation}
{\cal L}=-\frac 14F^{\mu \nu }F_{\mu \nu }+\frac 12D^\mu \phi D_\mu \phi
^{*}-U\left( \phi \right)  \label{eq1}
\end{equation}
in $1+2$ dimensions with a Minkowski space-time signature $\eta _{\mu \nu
}=(+,-,-)$. Here, $\phi$ is a complex scalar minimally coupled to an Abelian
gauge field, $U$ is an unknown potential and the covariant derivative is
defined as
\begin{equation}
D_\mu \phi =(\partial _\mu -ieA_\mu )\phi  \label{eq2}
\end{equation}

The equation of motion for $A_\mu $ is given by
\begin{equation}
\partial _\mu F^{\mu \nu }=J^\nu  \label{eq3}
\end{equation}
where $J^\nu $ is the N\"{o}ether's (conserved) current
\begin{equation}
J_\mu =-\frac{ie}2(\phi ^{*}D_\mu \phi -\phi D_\mu \phi )  \label{eq4}
\end{equation}

The energy momentum tensor obtained from (\ref{eq1}) is
\begin{equation}
T_{\mu \nu } =\frac 12(F_{\mu \alpha }F_{\,\,\nu }^\alpha +F_{\nu \alpha
}F_{\,\,\mu }^\alpha )+ \frac 12(D_\mu \phi ^{*}D_\nu \phi +D_v\phi
^{*}D_\mu\phi ) -\eta _{\mu \nu }{\cal L}  \label{eq5}
\end{equation}
and integration of the $T_{00}$ component yields
\begin{equation}
{\cal E}=\int d^2x\left\{ \frac 12(B^2+{\bf E}^2)+\frac 12\left| D_0\phi
\right| ^2+\frac 12\left| D_i\phi \right| ^2+U\right\}  \label{eq6}
\end{equation}

In order to fix the self-duality conditions, we will focus on static
topologically nontrivial solutions. Such classical configurations have been
shown to exist in this theory and are known as the Nielsen-Olesen vortices
\cite{Nielsen-Olesen}. By choosing the radiation gauge $A_0=0$ it can be
easily seen that these are electrically neutral solutions. This implies a
zero $J_0$ component and a vanishing electric field in the whole space.

Making use of the relation
\begin{equation}
\frac 12D_i\phi ^{*}D_i\phi =\frac 12|(D_1\pm iD_2)\phi |^2\pm \frac
1{2e}\varepsilon _{ij}\partial _iJ_j\mp \frac e2B|\phi |^2  \label{eq6.1}
\end{equation}
in eq.(\ref{eq6}), we obtain after some algebra
\begin{eqnarray}
{\cal E} &=&\frac{ev^2}2|\Phi _B|\pm \frac 1{2e}\oint_{r\rightarrow \infty }%
{\bf J.}d{\bf l\pm }\int d^2x\left\{ \frac 12\left( B\pm \sqrt{2U}\right)
^2\right.  \nonumber \\
&&\ \left. \pm \left[ \frac e2\left( v^2-|\phi |^2\right) -\sqrt{2U}\right]
B+\frac 12|(D_1\pm iD_2)\phi |^2\right\}  \label{eq9}
\end{eqnarray}
where the upper (lower) sign corresponds to positive (negative) values of
the magnetic flux $\Phi _B\equiv -\int d^2xB$. Let us remind that $\Phi _B$
yields topological nontrivial classes for a potential allowing asymmetric
vacuum phases.

Written in this form, the duality conditions to be imposed become apparent,
\begin{eqnarray}
D_1\phi &=&\mp iD_2\phi  \nonumber \\
B &=&\pm \frac e2\left( |\phi |^2-v^2\right)  \label{eq9.22}
\end{eqnarray}
Note that since we are considering only finite energy configurations, the
line integral of the vector current vanishes because the spatial components
of covariant derivatives have to be zero at spatial infinity.

In the Bogomol'nyi limit, given by eq.(\ref{eq9}), it can be seen that a
lower bound for the energy exists; namely ${\cal E}\geq ev^2|\Phi _B|/2$,
where the equal sign is saturated for
\begin{equation}
U(|\phi |^2)=\frac{e^2}8\left( |\phi |^2-v^2\right) ^2,  \label{eq9.2}
\end{equation}
which is a well-known result.


\subsection*{Chern-Simons-Higgs Model}

Bogomol'nyi-type vortex solutions are also en\-coun\-tered in the CSH model,
whe\-reas in this case vortices are electrically charged and the Higgs
potential is of sixth-order \cite{JW,JLW,Hong-Kim-Pac}. On the other side,
nontopological soliton solutions are also supported but in this case their
magnetic fluxes are not quantized \cite{JLW}.

In order to explicitly obtain this potential let us define the Lagrangian
density
\begin{equation}
{\cal L}=\frac \kappa 4\varepsilon ^{\mu \nu \rho }A_\mu F_{\nu \rho }
+\frac 12D^\mu \phi D_\mu \phi ^{*}-U(\phi).  \label{eq10}
\end{equation}
Now, the equation of motion for the gauge field is a first order one, being
given by
\begin{equation}
\kappa F^\mu =J^\mu  \label{eq11}
\end{equation}
where $F^\mu \equiv \frac 12$ $\varepsilon ^{\mu \nu \rho }F_{\nu \rho }$
and $J^\mu \,$ is the same as in (\ref{eq4}). The energy momentum tensor is
derived as usual
\begin{equation}
T_{\mu \nu }=\frac 12(D_\mu \phi ^{*}D_\nu \phi +D_v\phi ^{*}D_\mu \phi
)+\eta _{\mu \nu }(\frac 12|D_\alpha \phi |^2+U)  \label{eq12}
\end{equation}
so that the energy follows. Note however that the CS term is absent, thus,
the energy functional simply reads
\begin{equation}
{\cal E}=\int d^2x\left\{ \frac 12D_0\phi ^{*}D_0\phi +U+\frac 12D_i\phi
^{*}D_i\phi \right\}  \label{eq13}
\end{equation}
Following similar steps as in the previous section, it can be written as
\begin{eqnarray}
{\cal E} &=&\frac{ev^2}2|\Phi_B| \pm \int d^2x\left\{ \frac 1{2\left| \phi
\right| ^2}\left| \phi ^{*}D_0\phi \mp i\sqrt{2U\left| \phi \right| ^2}%
\right| ^2\pm \frac{\sqrt{2U} }e\frac{J_0}{\left| \phi \right| }\right.
\nonumber \\
&&\left. +\frac 12|(D_1\pm iD_2)\phi |^2\mp \frac e2B(|\phi|^2-v^2)\right\}
\label{eq13.2}
\end{eqnarray}
Notice that the second term in the equation above forces the magnetic field $%
B$ $(=-J_0/\kappa)$ to vanish wherever $\phi$ does, so that the vortex
becomes closed and the field lies within a toroidal region. This term,
causing this kind of ring vortex configurations is physically relevant in
order to stabilize nontopological soliton solutions. Compare it to $%
\kappa^2B^2/4e^2|\phi |^2$, obtained in ref.\cite{JLW} for a particular
sixth order Higgs potential (actually, this expression is
reobtained when a lowest energy configuration is considered (see
below)).

By imposing the following self-duality conditions
\begin{eqnarray}
\phi ^{*}D_0\phi &=&\pm i\sqrt{2|\phi |^2U}  \nonumber \\
D_1\phi &=&\mp iD_2\phi  \label{eq14}
\end{eqnarray}
and using the time component of eq.(\ref{eq11}) one has
\begin{equation}
{\cal E}=\frac e2v^2|\Phi _B|\pm \int d^2x\left( \frac e2(v^2-|\phi |^2)-%
\sqrt{\frac{2U}{|\phi |^2}}\kappa \right) B.  \label{eq15}
\end{equation}
Again, for
\begin{equation}
U(|\phi |^2)=\frac{e^4}{8\kappa ^2}|\phi |^2\left( |\phi |^2-v^2\right) ^2.
\label{eq16}
\end{equation}
the lower bound is reached. Now, equation (\ref{eq14}) can be manipulated so
as to produce a self-duality condition in terms of $J_0$%
\begin{equation}
J_0=\pm e\sqrt{2|\phi |^2U}  \label{eq16.2}
\end{equation}
Hence, using once more eq.(\ref{eq11}), eq.(\ref{eq16}) implies
\begin{equation}
B=\pm \frac{e^3}{2\kappa ^2}|\phi |^2\left( |\phi |^2-v^2\right)
\label{eq17}
\end{equation}
in agreement with the literature \cite{JW,JLW}. Notice that in this
(minimal) model a constraint on the gauge field arise; namely, $A_0$ is
related to the magnetic field by means of the time component of eq. (\ref
{eq11}), {\it i.e.},
\begin{equation}
A_0=\frac e{\kappa ^2}\frac B{\left| \phi \right| ^2}.
\end{equation}
This implies of course that the radiation gauge choice is prohibited in this
case.


\subsection*{Maxwell-Chern-Simons Higgs Model}

In the present section we would like to recover the well known Higgs
potential for the spontaneously broken Maxwell-Chern-Simons model \cite{LLM}%
. The (minimally coupled) Lagrangian density reads
\begin{eqnarray}
{\cal L} &=&-\frac 14F^{\mu \nu }F_{\mu \nu }+\frac \kappa 4\varepsilon
^{\mu \nu \rho }A_\mu F_{\nu \rho }+\frac 12D^\mu \phi D_\mu \phi ^{*}
\nonumber \\
&&+\frac 12\partial _\mu N\partial ^\mu N-V-|\phi |^2W  \label{eq19}
\end{eqnarray}
where\ for convenience, the potential has been written as
$U=V+|\phi |^2W$ and $N$ is a neutral scalar field.

The equation of motion for the gauge field now reads
\begin{equation}
\partial _\mu F^{\mu \nu }+\kappa F^\nu =J^\nu  \label{eq20}
\end{equation}
implying the following ``Gauss Law''
\begin{equation}
\partial _iE_i+\kappa B+J_0=0  \label{eq21}
\end{equation}
On the other side, for a static $N$ field the energy functional of the model
is given by
\begin{eqnarray}
{\cal E} &=&\int d^2x\left\{ \frac 12(B^2+{\bf E}^2)+\frac 12D_0\phi
^{*}D_0\phi +\frac 12D_i\phi ^{*}D_i\phi \right.  \nonumber \\
&&\ \ \ \left. \;\;\;\;+\frac 12\partial _iN\partial _iN+V(|\phi |^2)+|\phi
|^2W\right\}  \label{eq22}
\end{eqnarray}
which after some algebra reads
\begin{eqnarray}
{\cal E} &=&\int d^2x\left\{ \frac 12(B\pm \sqrt{2V})^2\mp B\sqrt{2V}\pm
\frac{\sqrt{2W}}eJ_0\right.  \nonumber \\
&&\ \ \;+\frac 12\left| D_0\phi \mp i\sqrt{2W}\phi \right| ^2+\frac
12|(D_1\pm iD_2)\phi |^2  \label{eq22.2} \\
&&\ \ \ \ \left. \mp \frac e2B|\phi |^2+\frac 12(E_i\pm \partial _iN)^2\mp
E_i\partial _iN\right\}  \nonumber
\end{eqnarray}
By imposing the following self-duality conditions
\begin{eqnarray}
B &=&\mp \sqrt{2V}  \nonumber \\
D_0\phi &=&\pm i\sqrt{2W}\phi  \nonumber \\
D_1\phi &=&\mp iD_2\phi  \label{eq23} \\
E_i &=&\mp \partial _iN\;,  \nonumber
\end{eqnarray}
eliminating $J_0$\thinspace (eq. \ref{eq21}), and integrating $E_i\partial
_iN$ by parts we have
\begin{eqnarray}
{\cal E} &=&\frac{ev^2}2|\Phi _B|+\int d^2x\left\{ \pm B\left[ \frac
e2(v^2-|\phi |^2)-\sqrt{2V}-\frac{\sqrt{2W}}e\kappa \right] \right.
\nonumber \\
&&\ \ \ \left. \pm \left( N-\frac{\sqrt{2W}}e\right) \partial _iE_i\right\}
\;  \label{eq24}
\end{eqnarray}
It is clear that the system will lie on its lower bound limit provided that
\begin{eqnarray}
\sqrt{2V} &=&\frac e2(v^2-|\phi |^2)-\frac{\sqrt{2W}}e\kappa  \nonumber \\
\frac{\sqrt{2W}}e &=&N  \label{eq25}
\end{eqnarray}
From eqs.(\ref{eq23}) and (\ref{eq25}) above, one can read out the relation
between the magnetic and the scalar fields
\begin{equation}
B=\pm \frac e2(|\phi |^2-v^2+\frac{2\kappa }eN)
\end{equation}
bringing about the explicit form of the Higgs potential
\begin{equation}
U(N,|\phi |^2)=\frac{e^2}2N^2|\phi |^2+\frac{e^2}8(|\phi |^2-v^2+\frac{%
2\kappa }eN)^2  \label{eq27}
\end{equation}
which coincides with the potential used in \cite{LLM}.


\section{Nonminimal Models}

In three space-time dimensions Pauli type coupling is known to give a
meaningful contribution to the magnetic moment of fields, without any
reference to their spin statistics. Namely, even scalar fields can present a
nonzero magnetic moment. The interest on such `nonminimal' coupling was
recently renewed \cite{Stern} since, using some critical value for this kind
of coupling in a MCSH model, one recovers the ideal anyon behavior proper of
pure CS theory. Later on further investigation was performed including a
nontopological Higgs potential \cite{Torres,LeeMin,Ghosh} where axially
symmetric self-dual solutions were encountered for a critical value of the
anomalous coupling.

Here we analyze the MCSH model in the topological sector and we relax the
condition on the nonminimal coupling out of its critical value. Further, we
do not impose rotational symmetry before setting the Bogomolnyi limit in
order to ensure minimal energy solutions.

\subsection*{Maxwell-Chern-Simons-Higgs model with anomalous magnetic moment}

Let us consider a MCSH Lagrangian with the anomalous magnetic moment term
characterized by the coupling constant $g$
\begin{equation}
{\cal L}=-\frac 14F^{\mu \nu }F_{\mu \nu }+\frac \kappa 4\varepsilon ^{\mu
\nu \rho }A_\mu F_{\nu \rho }+\frac 12\nabla ^\mu \phi \nabla _\mu \phi
^{*}-U(\phi )  \label{eq50}
\end{equation}
with $\nabla _\mu $ defined as
\[
\nabla _\mu \phi \equiv (\partial _\mu -ieA_\mu -igF_\mu )\phi
\]
The equation of motion for $A_\mu \,$ is
\begin{equation}
\partial _\mu F^{\mu \rho }+\kappa F^\rho ={\cal J}^\rho +\frac
ge\varepsilon ^{\mu \nu \rho }\partial _\mu {\cal J}_\nu  \label{eq51}
\end{equation}
where ${\cal J}_\mu $ is defined by
\[
{\cal J}_\mu \equiv -\frac{ie}2(\phi ^{*}\nabla _\mu \phi -\phi \nabla _\mu
\phi )
\]
The time component of eq.(\ref{eq51}) defines the modified ``Gauss Law''
\begin{equation}
\partial _iE_i+\kappa B+\frac ge\varepsilon _{ij}\partial _i{\cal J}_j+{\cal %
J}_0=0  \label{eq53}
\end{equation}
The gauge invariant modes are now short-range due the mass term resulting
from the modified equation of motion. Hence, the first term in eq.(\ref{eq53}%
) has a vanishing integral. On the other hand, the third term results in a
line integral taken at infinity which also vanishes for finite energy
configurations. Therefore, it can be seen from the remaining piece that one
has the charge of the vortex solutions related to nonzero magnetic fluxes as
follows
\begin{equation}
Q=\kappa \Phi _B  \label{Q}
\end{equation}
Unlike the preceding models, the nonminimal coupling allows a
temporal gauge choice $A_0=0$, which simplifies the handling of the
equations. For example, the electric charge reads
\begin{equation}
Q=\int d^2x\ J_0=e\,g\int d^2x\ |\phi |^2B  \label{QQ}
\end{equation}
Then, from eqs.(\ref{Q},\ref{QQ}) (assuming that $-ge/\kappa >0$) it follows
that
\begin{equation}
-gev^2/\kappa >1  \label{rel1}
\end{equation}
where $v$ gives the minimum value of the symmetry breaking potential. (Of
course it is true provided that $v$ is the maximum value of the field $\phi $%
; it will be shown below.)

On the other hand, the energy functional is given by
\begin{equation}
{\cal E}=\int d^2x\left\{ \frac 12G({\bf E}^2+B^2)+\frac 12D_0\phi
^{*}D_0\phi +\frac 12D_i\phi ^{*}D_i\phi +U\right\}  \label{eq54}
\end{equation}
with
\[
G=1-g^2|\phi |^2
\]
In order to ensure a positive definite energy one then needs
\begin{equation}
|g|<1/v  \label{rel2}
\end{equation}
which together with eq.(\ref{rel1}) implies
\begin{equation}
\kappa <ev
\end{equation}
and therefore
\begin{equation}
|g|<e/\kappa  \label{rel3}
\end{equation}
Actually, this relation excludes the constraint usually imposed on $g$ in
order to obtain a set of equations of motion of the first order without
imposing self-duality conditions \cite{Torres,LeeMin}.

In the $A_0=0$ gauge, and after some calculation, the energy functional can
be written as
\begin{eqnarray}
{\cal E} &=&\frac{ev^2}2|\Phi _B|+\int d^2x\left\{ \pm \left[ \frac
e2(v^2-|\phi |^2)-\sqrt{2GU}\right] B\right.   \nonumber \\
&&\ \ \left. \frac 12G\left( B\pm \sqrt{2UG}\right) ^2+\frac 12\left|
(D_1\pm iD_2)\phi \right| ^2\right\}   \label{eq55}
\end{eqnarray}
By imposing the self-dual equations,
\begin{eqnarray}
D_1\phi  &=&\mp iD_2\phi   \nonumber \\
B &=&\pm G^{-1}\frac e2(|\phi |^2-v^2)  \label{eq56}
\end{eqnarray}
the {\it topological} potential can be then determined in order to
achieve the lower bound limit. So we have,
\begin{equation}
U(|\phi |^2)=G^{-1}\frac{e^2}8(|\phi |^2-v^2)^2 . \label{dualeq}
\end{equation}
As already pointed out in the introduction, this potential leads to
topologically stable vortex solutions for the Maxwell-Chern-Simons
Higgs model with no reference to scalar field introduced {\it
ad-hoc} \cite{LLM}.

Now, for small values of $g$ we can perform a
series expansion of the topological Higgs potential just found, in
order to obtain the non anomalous phase limit
\begin{equation}
U(|\phi |^2)=\frac{e^2}8(|\phi |^2-v^2)^2+g^2\frac{e^2}8|\phi |^2(|\phi
|^2-v^2)^2+\dots   \label{serie1}
\end{equation}
Notice that in the $g=0$ limit the sixth order term, characteristic of a CS
contribution, is absent. This is consistent with eqs.(\ref{Q}-\ref{rel1})
which impose a vanishing $\kappa $ whenever $g\rightarrow 0$.

Let us write eq.(\ref{serie1}) in the following way
\begin{equation}
U(|\phi |^2)=\frac{e^2}8(|\phi |^2-v^2)^2+m^2\frac{e^4}{8\kappa ^2}|\phi
|^2(|\phi |^2-v^2)^2+\dots  \label{serie2}
\end{equation}
with $m<1$, defined by $|g|=me/\kappa $ [see eq.(\ref{rel3})] (notice that
while eq.(\ref{rel1}) imposes that $\kappa /ev^2<|g|$,\ $m$ is assumed to be
small enough to make the CS term above a second order correction). Thus, for
a small a.m.m one can see in eq.(\ref{serie2}) the remaining of the more
elementary theories contained in the present model, namely, both a Maxwell
(Nielsen-Olesen) topological potential term and a typical sixth order CS one.

The theory possess two massive gauge propagating modes. The masses of these
gauge excitations read
\begin{equation}
m_{A\pm }=\frac{\kappa I}{2G_v}\pm \sqrt{(\frac{\kappa
I}{2G_v})^2+\frac{e^2v^2}{G_v}}
\end{equation}
where $I=1+4egv^2/\kappa $ and $G_v=1-g^2v^2$, and for
$g\rightarrow 0$ one obtains the Nielsen-Olesen model mass
$m_{NO}=ev$. On the other hand, the Higgs mass is easily seen to be
\begin{equation}
m_H={ev}/{\sqrt{G_v}}
\end{equation}
which of course approaches also $m_{NO}$ for a vanishing $g$.


\subsection*{Solutions and numerical analysis}

\vskip .3cm

By means of proper gauge transformations, static rotationally symmetric
configurations can be written in polar coordinates as
\begin{eqnarray}
&&\phi (r)=vR(r)e^{in\theta } \\
&&e{\bf A}(r)=-\frac{\hat \theta }r\left[ a(r)-n\right]
\end{eqnarray}
where $R$ and $a$ are real functions of $r$, and $n$ an integer
indicating the topological charge of the vortex. Now, under
transformation $ r\rightarrow (\sqrt{2}/ev) \, r$ the self-duality
equations (\ref{eq56}) become
\begin{eqnarray}
&&R^{\prime }=\pm \frac a r\ R \\
&&\frac{a^{\prime}}r =\pm \frac 1{1-{\gamma }^2R^2}(R^2-1)
\end{eqnarray}
where $\gamma <1$ is defined by $g=\gamma /v$.

The natural boundary conditions at spatial infinity result from the
requirement of finite energy, namely, $R(\infty )=1$ and $a(\infty )=0$ for
any nontrivial vorticity $n$. On the other hand, at the origin one must
expect nonsingular fields, implying $R(0)=0$ and $a(0)=n$. Hence, the
magnetic field reads
\begin{equation}
B=-\frac{ev^2}{2}\ \frac{a^{\prime}}r  \label{Bfield}
\end{equation}
and its flux is, as expected
\begin{equation}
\Phi _B=\frac{2\pi }e[a(0)-a(\infty )]=\frac{2\pi}en.  \label{fluxB}
\end{equation}
At large values of $r$ it is easy to see that the $n>0$ solutions behave
like
\begin{equation}
R(r)\rightarrow 1-cK_0(r)
\end{equation}
and
\begin{equation}
a(r)\rightarrow d\, r K_1(r)
\end{equation}
where $c$ and $d$ are constants. The $n<0$ configurations are related to
these ones by the transformation $a\rightarrow -a$ and $R\rightarrow R$. The
behavior of the solutions at small values of $r$ is power like, and without
any loss of generality we can simply assume
\begin{eqnarray}
R(r) &\sim &c_nr,  \nonumber \\
\ \ \ \ \ a(r) &\sim &n
\end{eqnarray}
Notice that $c_n$ is determined by the shape of the fields at
infinity rather than their behavior at the origin. Indeed, we have
numerically solved the self-duality equations of motion by means of
an iterative procedure which involves a tentative value for $c_n$
which is corrected each time by imposing that both $R\rightarrow 1$
and $a\rightarrow 0$ hold together at infinity.

For $n=1,2$ and 3 we have found $c_1=8.891\times 10^{-1},$ $c_2=4.796\times
10^{-6}$ and $c_3 = 6.877\times 10^{-9}$ (see fig.1 for higher precision).
In fig.1 we show the topologically nontrivial solutions $R(r)$ and $a(r)$
and in fig.2 we plot the corresponding magnetic fields. In fig.3 we plot the
magnetic field for $n=1$ vorticity at different values of $\gamma$.


\unitlength=1cm
\begin{figure}[ht]
\centering
\begin{picture}(12,8)
\epsfig{file=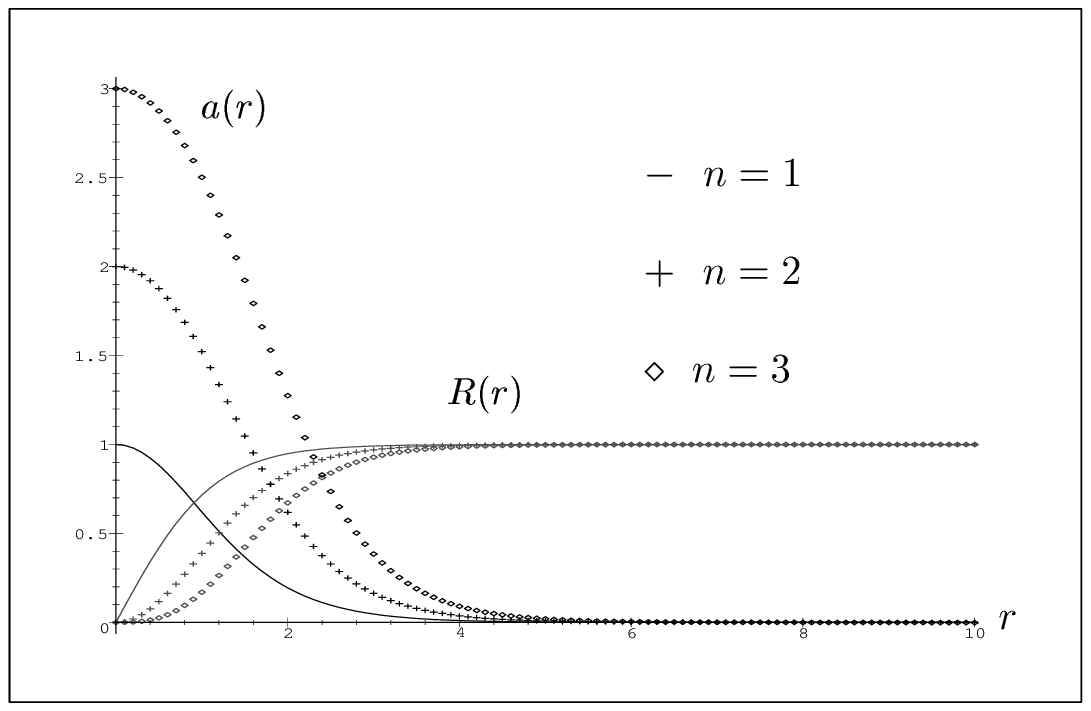,width=12cm,height=8cm}
\end{picture}
\caption{The scalar $R$ and the gauge field $a$ as a function of $r$. The
values of the $c_n$ constants are fixed by the shape of the fields at
infinity: (--)~$c_1~=~8.891308075\times 10^{-1}$, ($+$) $c_2=4.796825890%
\times 10^{-6}$, ($\diamond$) $c_3=6.877604870\times 10^{-9}$. }
\vskip 2cm

\begin{picture}(12,8)
\epsfig{file=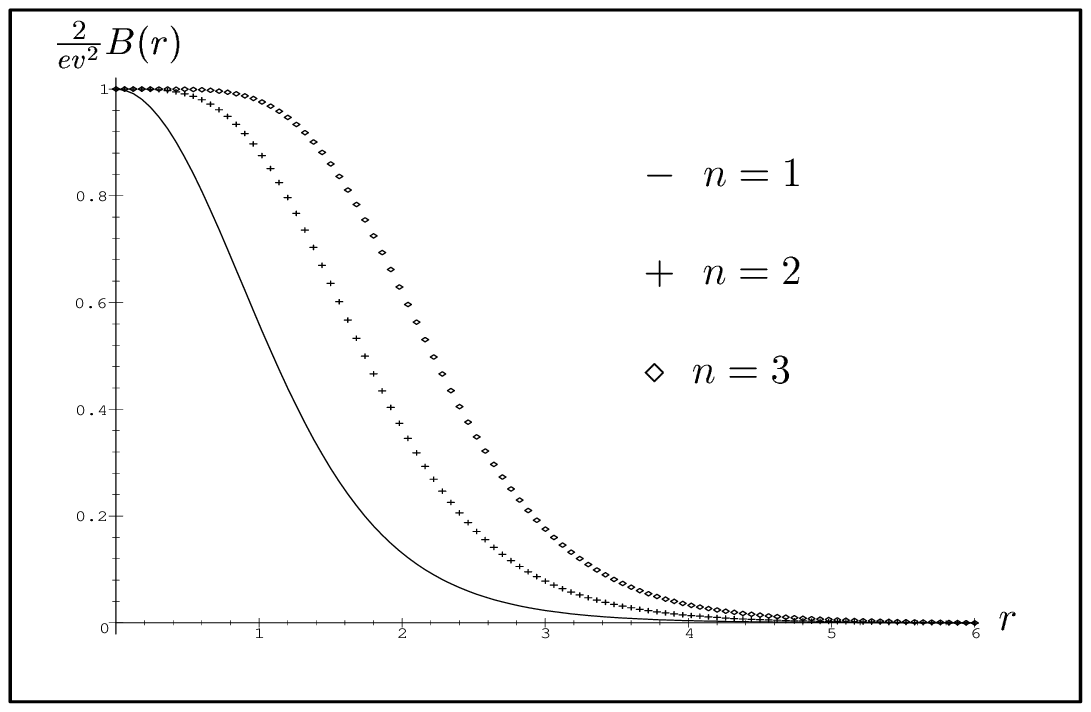,width=12cm,height=8cm}
\end{picture}
\caption{The magnetic field $B$ as function of $r$ for $n=$~1, 2, and 3
vorticities. Notice that the field distributions are concentrated at the
origin -- like NO vortices.}
\end{figure}

\newpage
\unitlength=1cm
\begin{figure}[ht]
\centering
\begin{picture}(12,8)
\epsfig{file=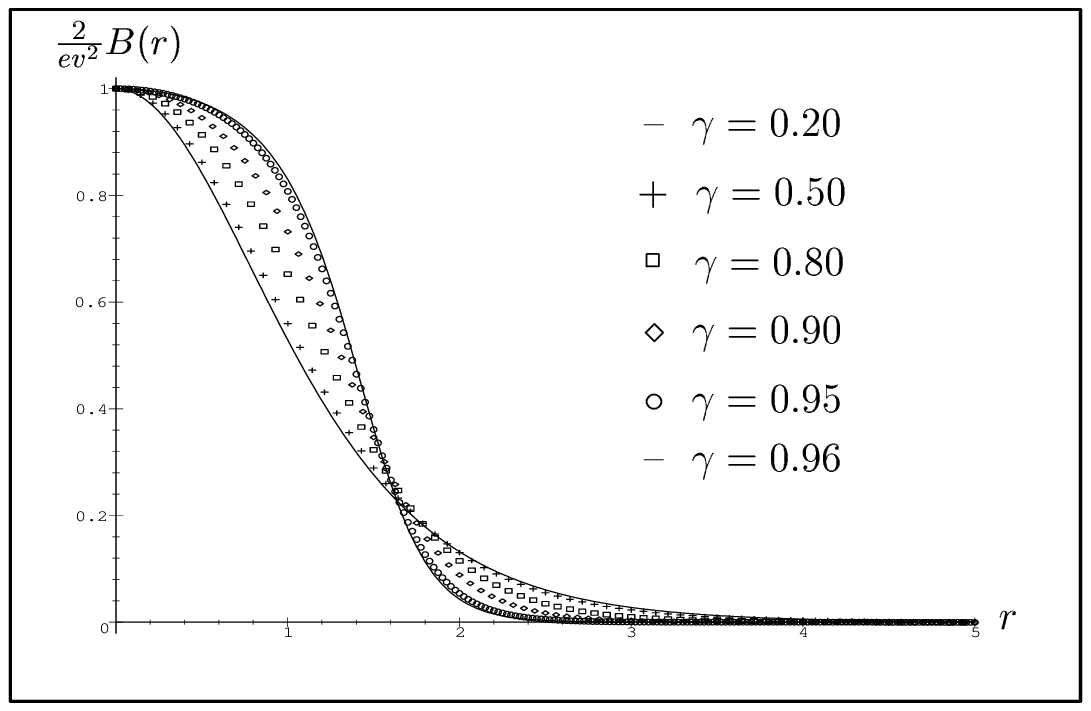,width=12cm,height=8cm}
\end{picture}
\caption{The magnetic field $B$ as a function of $r$ for various values of
the anomalous magnetic moment parameter $\gamma\, (=vg).$}
\end{figure}

\vskip 0.5cm \newpage


\section{Summary and conclusions}

We have shown that a Higgs potential, necessary to get vortex solutions of
minimal energy in the Bogomol'nyi limit, can be obtained by imposing
self-dual equations from the very beginning, in contrast to the regular
procedure found in the literature \cite{Nielsen-Olesen}-\cite{Torres}. From
this point of view, we have revisited the Abelian Higgs, Chern-Simons Higgs
and Maxwell-Chern-Simons Higgs models, and we have regained the standard
outcomes.

Thereafter, guided by this outlook, we obtained a topological Higgs
potential for a generalized MCSH theory, modified by the inclusion of a
nonminimal coupling controlled by a parameter $g$ which is introduced in the
covariant derivative. By means of a series expansion in $g$, this potential
has shown to contain traces of more elementary models, namely, Abelian Higgs
and Chern-Simons Higgs contributions, precisely in their usual topological
phase.

In order to get the topological potential, we worked in a temporal
gauge which clarifies the relations among the constants of the
model $e, \kappa, v$ and $g$. This gauge choice also prevents the
choice of the critical value $g=-e/\kappa $ commonly adopted; this
relation, in fact, would lead to the already known {\it
nontopological} solutions to the model \cite{Torres}%
.

The generalized MCSH model that we have analyzed, possess two
massive propagating modes. The masses are different from the Higgs
mass, even in the Bogomol'nyi limit, a result which is produced by
the anomalous magnetic moment. On the other hand, it has been
recently argued that bosonic theories in the Bogomol'nyi limit
could be closely connected to their $N=2$ ~supersymmetric extension
\cite{edels}. Thus, it is worth enquiring at this point if the
supersymmetric extension of this specific model requires different
conditions on the coupling constants (in this direction, see
\cite{navratil}). We hope to report on these issues in a future
work.

\section*{Acknowledgments}

The authors would like to thank Professor J.A. Helayel-Neto for
stimulating and useful discussions. M.S.C. would like to thank the
Physics Department of the Universidade Federal do Cear\'{a} for
warm hospitality. H.R.C. is grateful to Centro Latinoamericano de
F\'\i sica (CLAF-CNPq) and CBPF-CNPq for warm hospitality and
financial support. M.S.C. and C.A.S.A are financially supported by
CNPq (Brazilian Research Agency). \ \ \

\end{document}